\documentstyle[preprint,12pt,aps,epsf]{revtex}
\tightenlines
\begin{document}

\title{RESONANT RAMAN SCATTERING IN NaV$_2$O$_5$ AS A PROBE OF ITS ELECTRONIC
STRUCTURE } 
\author{M.J.Konstantinovi\'c $^{a}$, Z.V.Popovi\'c $^{b}$, T.Ruf
$^{a}$, M.Cardona $^{a}$, A.N.Vasil'ev$^{c}$, M.Isobe$^{d}$ and Y.Ueda$^{d}$}
\address{$^a$ Max-Planck-Institut f\"ur Festk\"orperforschung,Heisenbergstr.
1,D-70569 Stuttgart, Germany} 
\address{ $^b$ VSM-lab., Department of Physics,
K.U.Leuven, Celestijnenlaan 200 D, Leuven 3001, Belgium} 
\address{ $^c$ Low
Temperature Physics Department, Moscow State University, 119899 Moscow, Russia }
\address{ $^d$ Institute for Solid State Physics, The University of Tokio,
7-22-1 Roppongi, Minato-ku, Tokio 106, Japan} 
\maketitle 
\begin{abstract} In order to investigate the origin of the phase transition
observed in NaV$_2$O$_5$, as well as its electronic structure, we have measured
Raman intensities as a function of the laser wavelength above and below the
phase transition temperature.  In the polarized Raman spectra at room
temperature we observe resonant enhancement of the 969 $cm^{-1}$ phonon mode
when the laser energy approaches 2.7 eV, presumably related to the (p-d)
electron hopping band, $O_3$($p_y$)-V($d_{xy}$), at 3.2 eV.  The 969 $cm^{-1}$
mode originates from the stretching vibrations along the c-axis involving the
V-$O_1$ bonds.  Since an ellipsometric determination of the dielectric function
$\epsilon_{cc}$ yields no structure in the 1.7 to 5.5 eV photon energy range, we
conclude that plane bonds couple strongly with the apical oxygens leading to a
large Raman efficiency.  In the low-temperature Raman spectra, almost all modes
that become active below the phase transition temperature T$_c$=34 K show
resonant behavior.  The most interesting ones, those at 66 and 106 $cm^{-1}$,
possibly of magnetic origin, exhibit a resonant intensity enhancement,
approximately by an order of magnitude, for laser photon energies around 1.85 eV
with respect to 2.43 eV.  This resonance effect may be associated with a weak
absorption band around 2 eV.  Finally, a destructive interference between the
resonant and the nonresonant contribution to the Raman scattering amplitude
(i.e.  an antiresonance) is found in the spectra for most of the (bb)
low-temperature modes.
\end{abstract}

\section{Introduction}

NaV$_2$O$_5$ has been reported to be the second inorganic spin-Peierls compound
\cite {a1}, following CuGeO$_3$, with a phase transition temperature T$_c$=34 K.
However, recent experimental findings suggest a different character for the
phase transition of NaV$_2$O$_5$ as compared to that of CuGeO$_3$ \cite {a2}.
NaV$_2$O$_5$ is believed to be a nearly perfect realization of a quarter-filled
low-dimensional system, with possible charge ordering below the critical
temperature.  Such a scenario is supported by structural analysis \cite {a3},
nuclear magnetic resonance experiments \cite {a4} and Raman scattering studies
\cite {a5}.  These results indicate that the V ions are equivalent at room
temperature and in a mixed-valent state, V$^{4.5+}$.  When lowering the
temperature below $T_c$ the V sites change their valence from uniform to
alternating V$^{4+}$ and V$^{5+}$ \cite {a4}.  Two different models for the
electronic ordering in the low-temperature phase have been discussed.  One is
described as in-line \cite {a6} with the V$^{4+}$ ions arranged along the legs
of the ladder ({\bf b}-axis) and the other one with the V$^{4+}$ ions arranged
in a zig-zag manner along the {\bf b}-axis \cite {a7,a8}.  The first one
requires an additional phase transition, like the ordinary spin-Peierls
transition, for an opening of the spin-gap.  For the second one it is argued
that a spin-gap appears due to the zig-zag ordering itself \cite {a8}.  However,
in both cases the driving mechanism for a particular charge ordering has not
been elucidated.  In the Raman spectra this transition manifests itself through
the appearance of new modes whose origin is still not fully understood \cite
{a9,a10,a11}.  In general, Raman scattering intensities are strongly correlated
with the electronic structure of solids.  Their determination as a function of
the laser photon energy yields information about the energies of interband
transitions and their orbital character as has been highlighted recently by
resonance Raman measurements on high- temperature superconductors \cite {a12}.
Having this in mind, we have measured Raman intensities as a function of the
laser wavelength, both above and below the phase transition temperature.  On the
basis of the observed effects we discuss the origin of the resonances and the
electronic structure of NaV$_2$O$_5$.

\section{Experiment}

The measurements were performed on single crystals, with a size of approximately
1x3x1 mm$^3$ along the {\bf a}, {\bf b}, and {\bf c} axis, respectively,
prepared as described in \cite {a1}.  As exciting source we used eight lines
from Ar$^+$ and Kr$^+$ ion lasers, with photon energies between 1.55 to 2.71 eV.
The laser beam, with an average power of 5 mW, was focused on the (001) surface
of the crystals.  The spectra were measured in back-scattering geometry using a
Dilor XY triple monochromator equipped with a CCD camera.  Comparable relative
Raman intensities were obtained by normalizing the measured spectra with respect
to the calibrated spectral emission density of an Ulbricht sphere system
(Opto-Electronik Reiche GmbH), thus taking care of the varying spectral response
of both spectrometer and detector over the large wavelength range investigated.

\section{High temperature spectra}

The room temperature Raman spectra in $c(aa)\bar c$ and $c(bb)\bar c$
 polarized configurations
are presented in Fig.  \ref{fig1}a) and b).  In the (aa) spectra seven phonon
modes are clearly seen with an additional broad continuum peaked at 650
$cm^{-1}$.  In this geometry all modes, as well as the continum, decrease in
intensity with decreasing laser photon energy.  In the spectra measured with
energies lower than 2 eV the appearance of an additional continuum centered at
about 170 $cm^{-1}$ is observed.  In contrast to the (aa) spectra, most peaks in
the (bb) spectra do not vary much with the laser line used, except for the 534
$cm^{-1}$ mode.  According to a lattice dynamical calculation \cite {a13}, the
969 $cm^{-1}$ mode originates from apical oxygen vibrations (V-$O_1$ stretching
vibrations) along the {\bf c}-axis.  The increase of its (aa) intensity with
increasing laser energy can be related to the resonance of the incoming (or
scattered) laser light with the strong electronic absorption band at 3.22 eV,
observed by ellipsometry \cite {a14}.  According to LMTO calculations \cite
{a15} this band represents the excitation of electrons hopping from the
$O_3$($p_y$) to the V($d_{xy}$) orbital.  Since the dielectric function along
the c-axis has no structure in the 1.7 to 5.5 eV range, we conclude that plane
bonds strongly couple with the apical oxygen leading to a large Stokes-Raman
efficiency, $S_{xx}(O_1)$.  Similarly, an electronic band related to
$O_3$($p_x$)-V($d_{xy}$) hopping is found along the {\bf b}-axis at 3.73 eV
\cite{a14}.  We thus expect a large intensity for the (bb) 969 $cm^{-1}$ phonon
at a laser energy around 3.7 eV.  One can use similar arguments for the resonant
behavior of other phonons with displacements that influence the O-V hopping.
The (aa) modes with energies of 448 and 534 $cm^{-1}$ overlap with the broad
continuum centered at 650 $cm^{-1}$ and exhibit typical Fano-type lineshapes
\cite {a16}.  We analyse the lineshape of the 535 $cm^{-1}$ mode by fitting to
the standard Fano expresion \cite {a16}.  The fitting parametar $q^{-1}$ is
plotted vs.  excitation energy in the inset of Fig.\ref{fig1}.  From the linear
extrapolation we obtain a resonance energy of $E_0$=0.9 0.2 eV.  This energy can
be associated with a large absorption band centered around 1-1.2 eV \cite
{a17,a18}, observed in both the {\bf E}$||${\bf a} and the {\bf E}$||${\bf b}
optical absorption spectra.  This band has been tentatively assigned to d-d
transitions \cite {a17,a18}.  In the literature there is still no consensus as
to whether the 1 eV band can be assigned to a d-d transition energy within the
t-J model using a reasonable range of parameters, with \cite {a18,a19} or
without \cite {a20} charge disproportionation.  An even more controversial issue
concerns the existence of low energy excitations in these models, as a possible
explanation for broad features found in IR and Raman spectra \cite {a12,a13}.

\section{Low temperature spectra}

The low-temperature (T=10 K) (aa) and (bb) Raman spectra are given in Fig.
\ref{fig2}a) and b).  We find new modes that become active below the phase
transition temperature, T$_c$=34 K, due to the doubling of the unit cell.  This
doubling can be explained by assuming the charge-ordering scenario with a
zig-zag ordering pattern of the charges \cite {a21}.  Almost all new modes show
resonant behavior.  The lowest frequency modes with energies at 66 and 106
$cm^{-1}$, previously assigned to magnetic excitations \cite {a10}, exhibit an
increase in intensity by approximately an order of magnitude when using a laser
photon energy around 1.85 eV (with respect to that at 2.43 eV) in (aa) polarized
spectra.  The intensity of the the 66 $cm^{-1}$ mode is plotted in Fig.
\ref{fig3}b) (aa - polarization) and Fig.  \ref{fig4}b) (bb - polarization) as a
function of the laser energy.  In the same figures we show $\epsilon_2^a$ and
$\epsilon_2^b$ for comparison, taken from \cite {a14}, respectively.  The peak
at 1.85 eV of the 66 $cm^{-1}$ mode intensity in (aa) polarization occurs close
to the broad structure centered around 2 eV .  The electronic transitions around
2 eV consist of two bands centered at 2 and 2.1 eV.  Their low absorption
strength, energy position and band separation of about 0.1 eV are typical for
crystal- field transitions of d-elements.  Thus, it is possible that these two
absorption bands correspond to crystal-field-excitations of $V^{4+}$ ions from
the ground xy state to $x^2-y^2$ and to $3z^2-r^2$ states.  The intensity
behavior of the 66 $cm^{-1}$ mode in (bb) polarization is quite different:
Intensity is increasing with increasing laser energy and has a local minimum
around 2.2 eV.  Again, such behavior is not easy to understand from simple
comparisom with $\epsilon_2$ and more correct analyses are necessary to clarify
these effects.  The difference in the resonance behavior of the 948 $cm^{-1}$
mode in the (aa) and the (bb) spectra, shown in Fig.  \ref{fig3}c) and Fig.
\ref{fig4}c), is also of interest.  While in the case of the (aa) spectrum this
mode looses intensity with decreasing laser energy, its behavior in the (bb)
spectra is exactly the opposite.  The origin of this effect is not clear.
However, it has been argued by Riera and Poilblanc \cite {a22} that a zig-zag
pattern of charges in a ladder can be stabilized by the Holstein coupling and
the nearest-neighbor Coulomb repulsion.  This indicates the importance of
electron-phonon interaction as a possible driving mechanism for charge-ordering
in NaV$_2$O$_5$.  If the 948 $cm^{-1}$ phonon is somehow involved in charge
ordering (via electron-phonon coupling) it would see such a particular
electronic structure \cite {a22} in its resonanace behaviour.  Finally, we
discuss the possible antiresonant behavior around 2.2 eV, suggested by
comparison of Fig.  \ref{fig4}a) and d).  Such an effect usually results from an
cancellation of resonant and nonresonant contributions to the scattering
amplitude.  Correspondingly, the cross section becomes very small below (or
above) the critical point energy.  However, it is also possible that there are
two resonances, one at 1.7 eV and one around 2.5 eV.  Thus phonons are resonant
either only at 1.7 eV (Fig.  \ref{fig3}b, \ref{fig4}c), only at 2.5 eV (Fig.
\ref{fig3}c), or at both energies (Fig.  \ref{fig4}b, d).  To resolve these
possibilities more acurate measurements are required.

\section{Conclusion}

In conclusion, we have presented measurements of
Raman intensities as a function of the laser wavelength above and below the
phase transition temperature.  We have found that various phonons show
significant resonances around 1.8 and 2.5 eV.  The first one can be probably
associated with weak crystal-field absorption bands around 2 eV, while the
second one may corresponds to the band which represents the excitations of
electrons hopping from the $O_3(p_y)$ to the $V(d_{xy})$ orbitals.

Acknowledgment: MJK thanks Roman Herzog-AvH foundation for financial support.

\begin{figure}
\caption 
{Room temperature Raman spectra of NaV$_2$O$_5$ single
crystals measured with several laser energies in a) $c(aa)\bar c$
 and b) $c(bb)\bar c$
scattering configurations.  L denotes plasma lines.  Inset: a plot of 
$q^{-1}$
versus excitation energy as obtained by fitting the line shape of the 
534 $cm^{-1}$ phonon.} 
\label{fig1} 
\end{figure}
\begin{figure}
\caption
{Raman spectra of NaV$_2$O$_5$ single crystals at T=10 K measured with
several laser energies in a) $c(aa)\bar c$ and b) $c(bb)\bar c$ 
scattering configurations.  
L denotes plasma lines.}
\label{fig2}
\end{figure} 
\begin{figure}
\caption 
{ a) $\epsilon_2^a$ at T=25 K, taken from [14]. 
 Intensity of b) 66 $cm^{-1}$
 mode and c) 948 $cm^{-1}$ mode as a function of laser energy (T=10 K). }
\label{fig3}
\end{figure} 
\begin{figure}
\caption
{ a) $\epsilon_2^b$ at T=25 K, taken from [14].
  Intensity of b) 66 cm-1, c) 948 cm-1
mode, d) 398 cm-1 mode as a function of laser energy (T=10 K).}
\label{fig4}
\end{figure} 

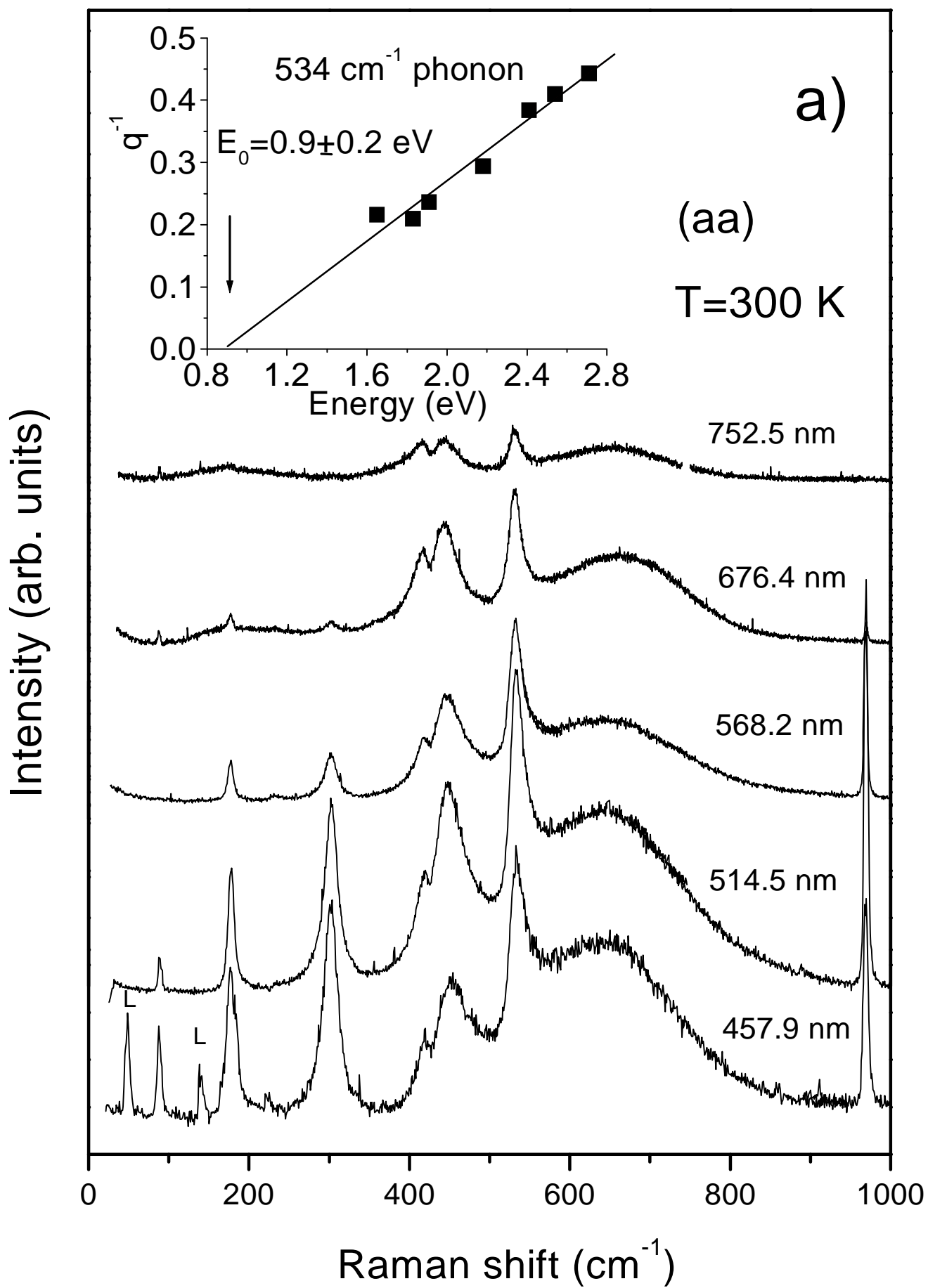
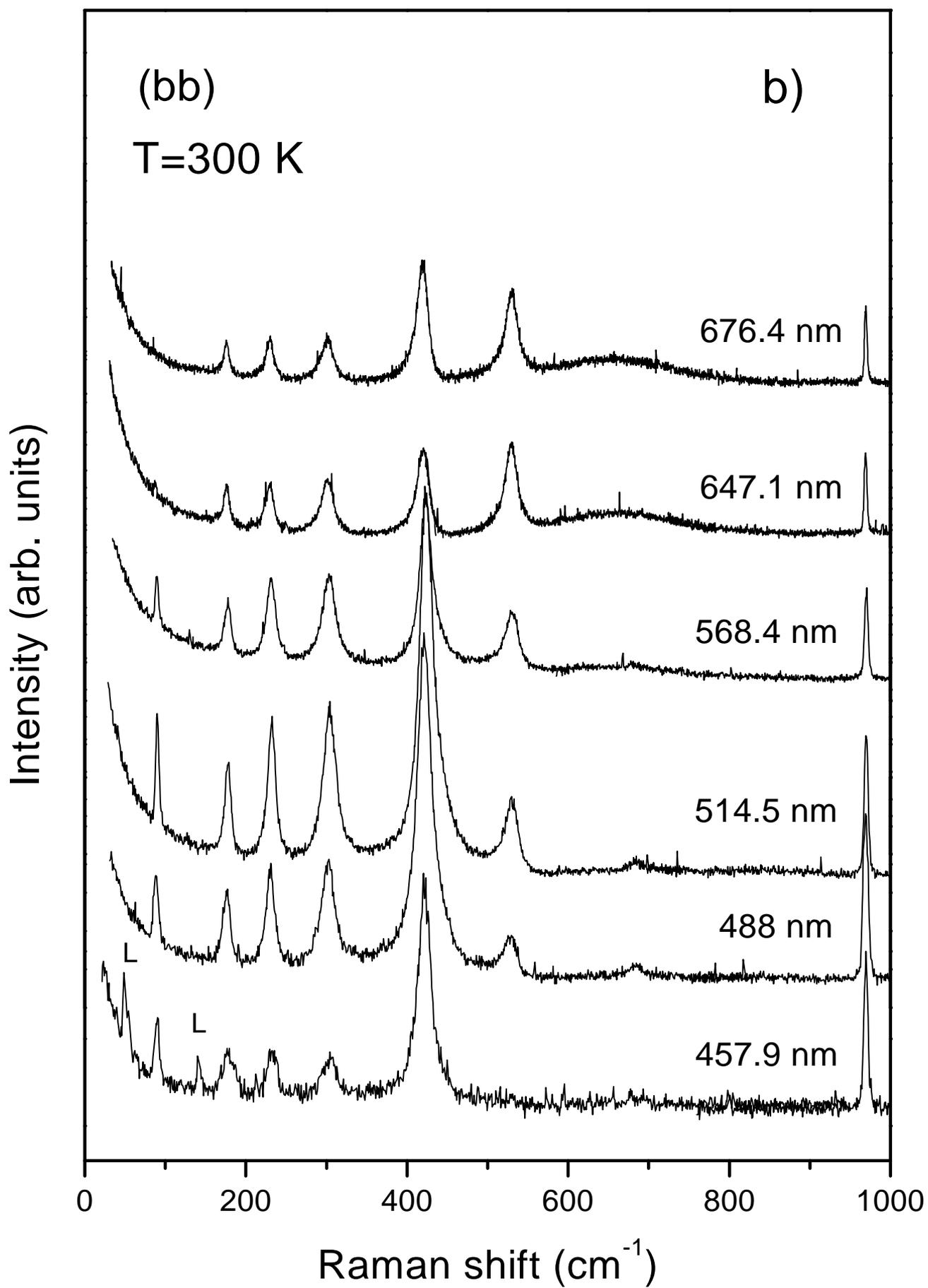
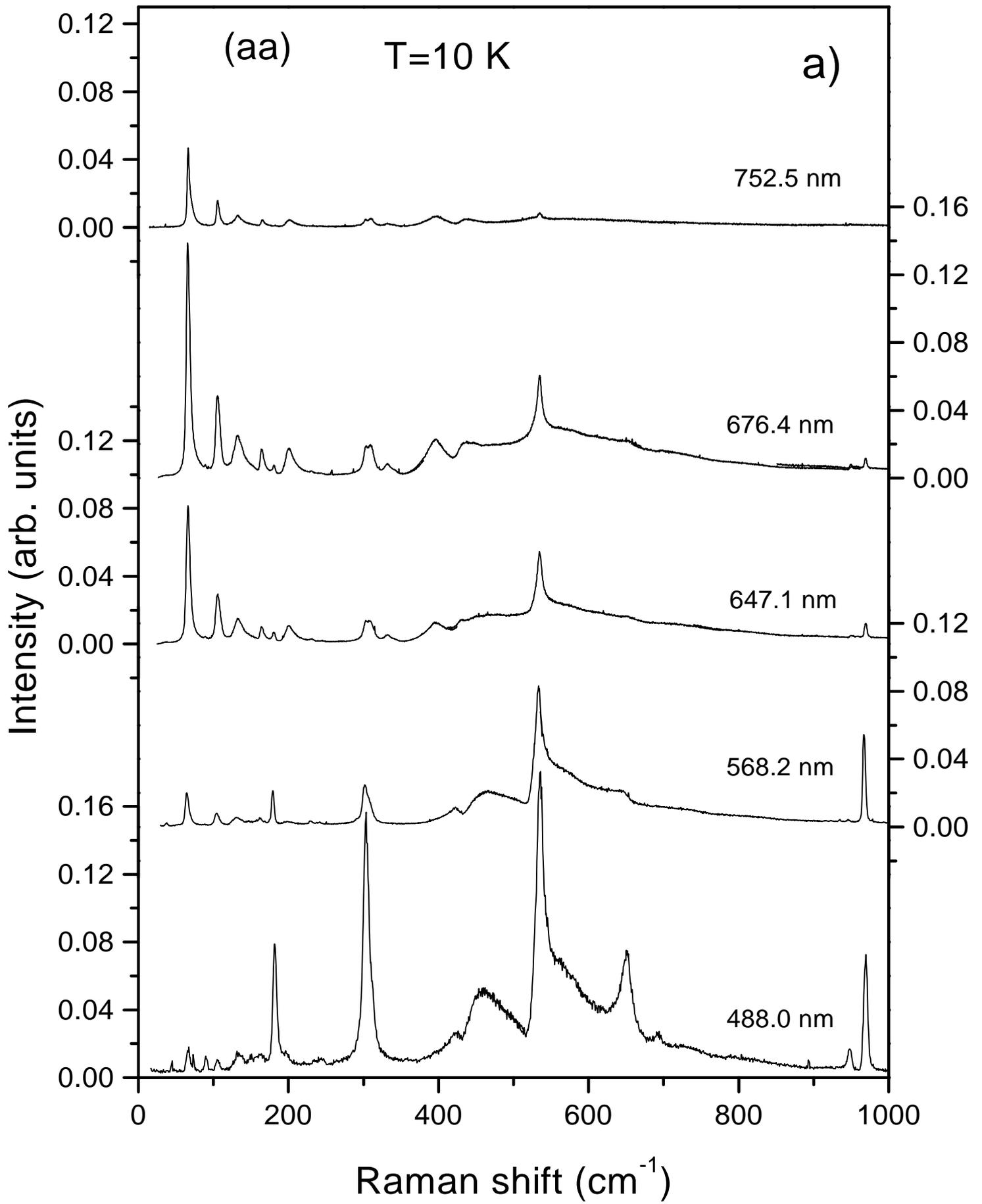
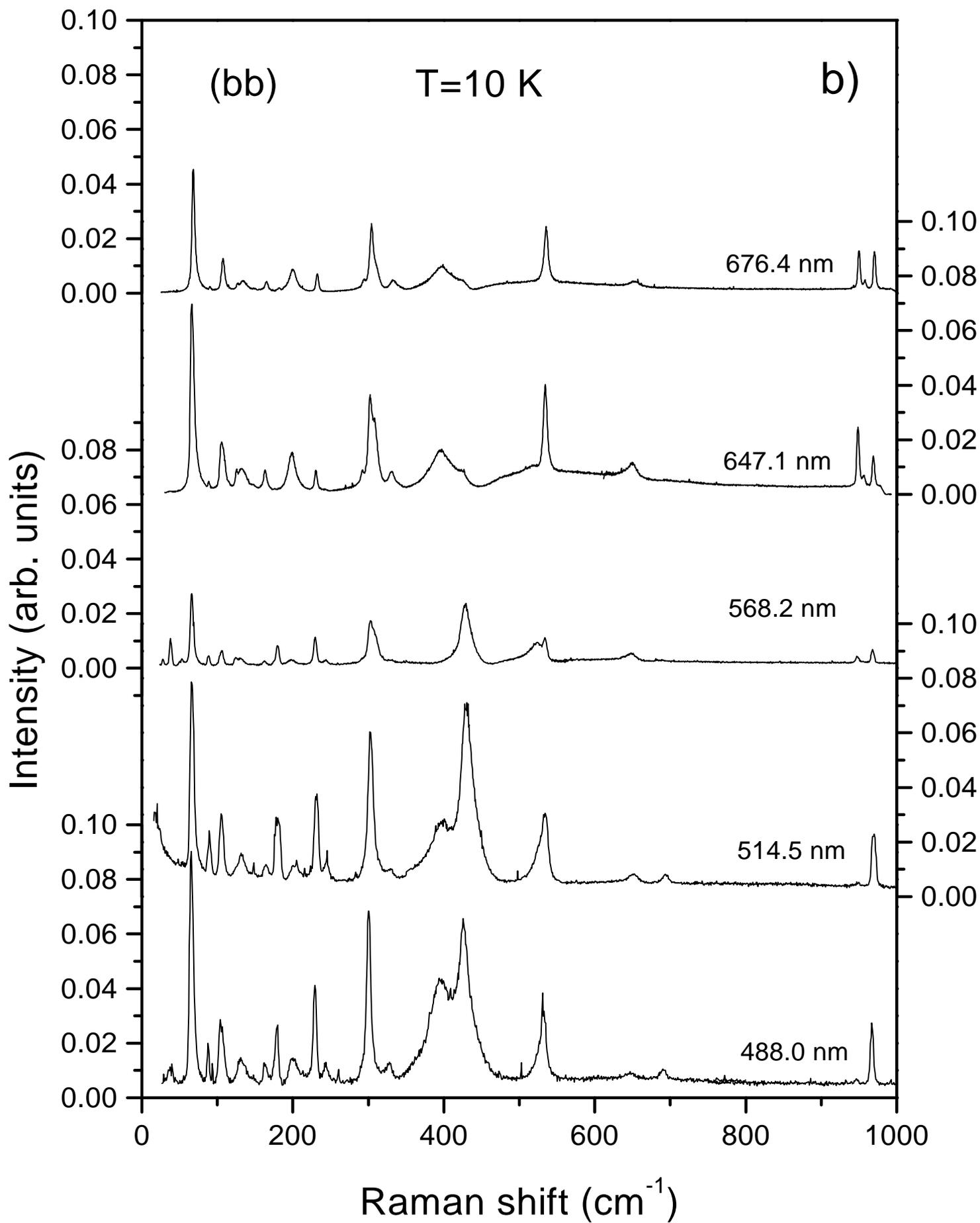
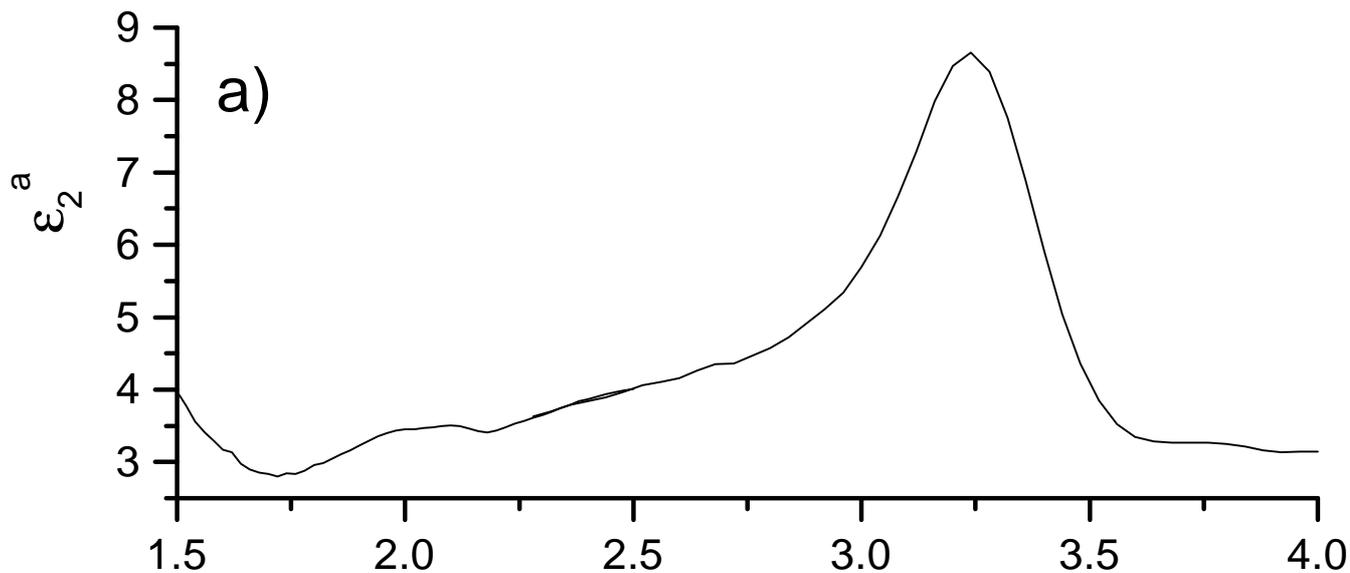
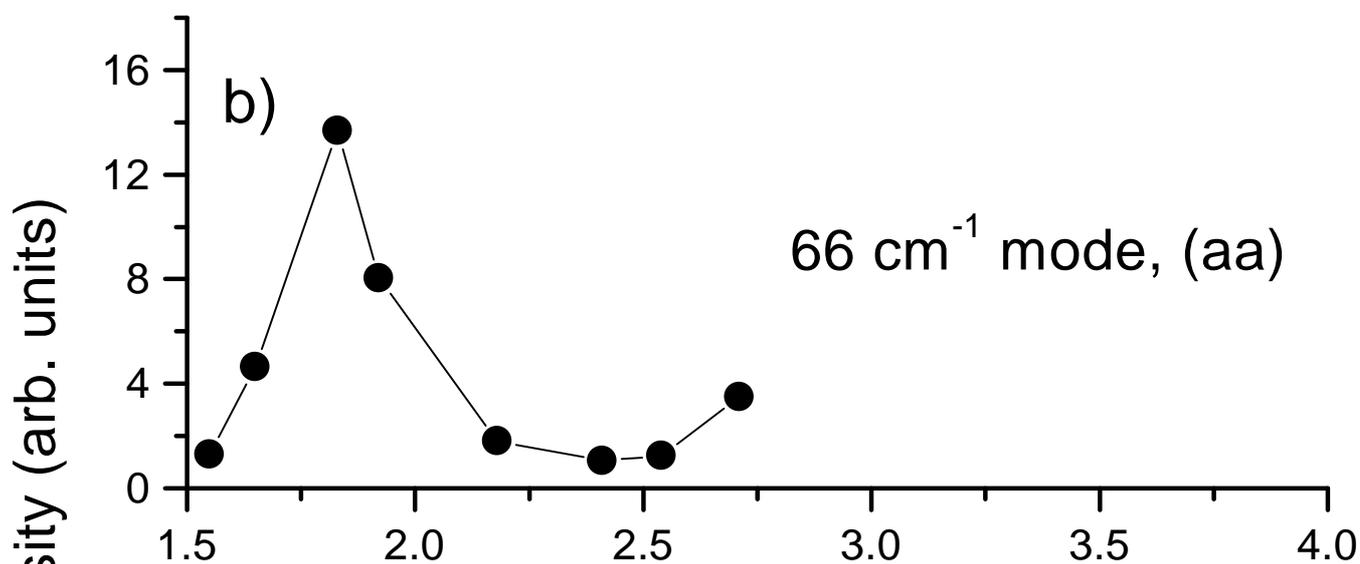
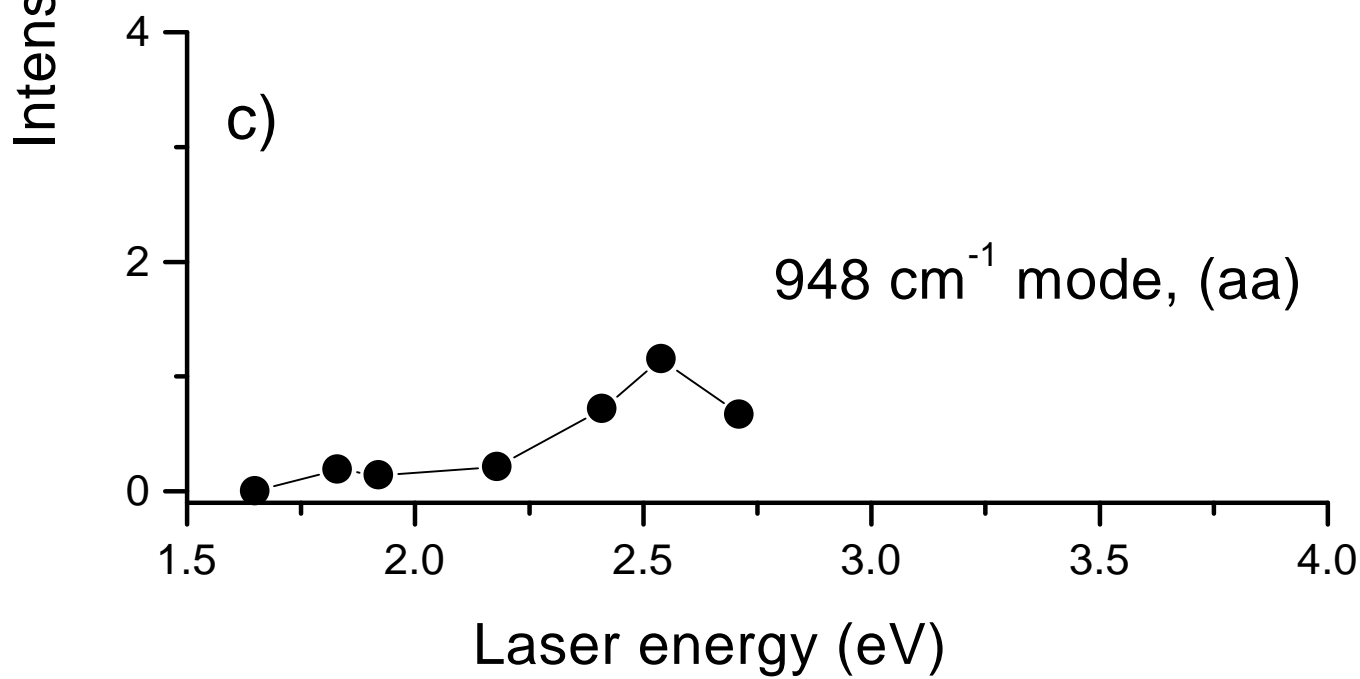
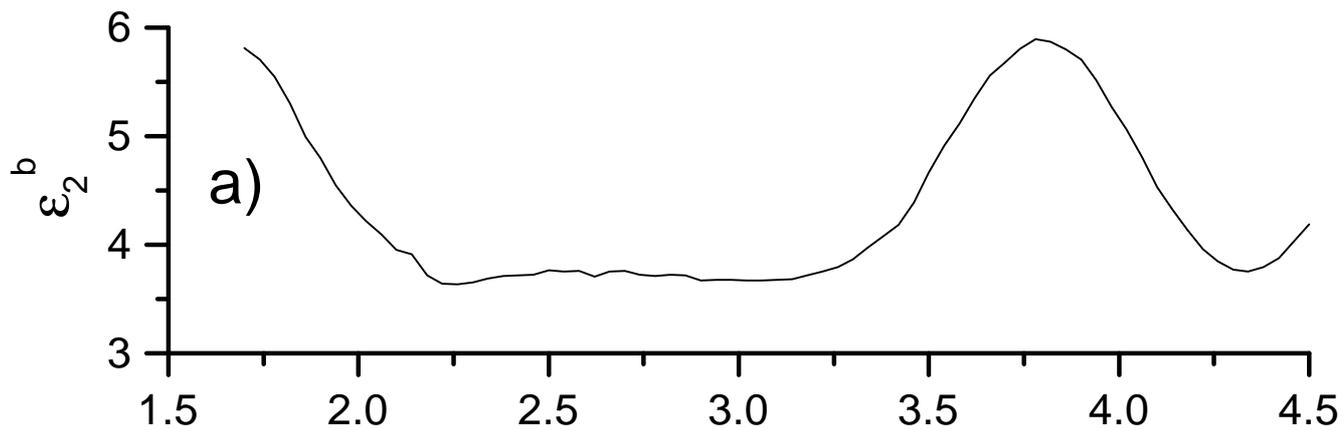
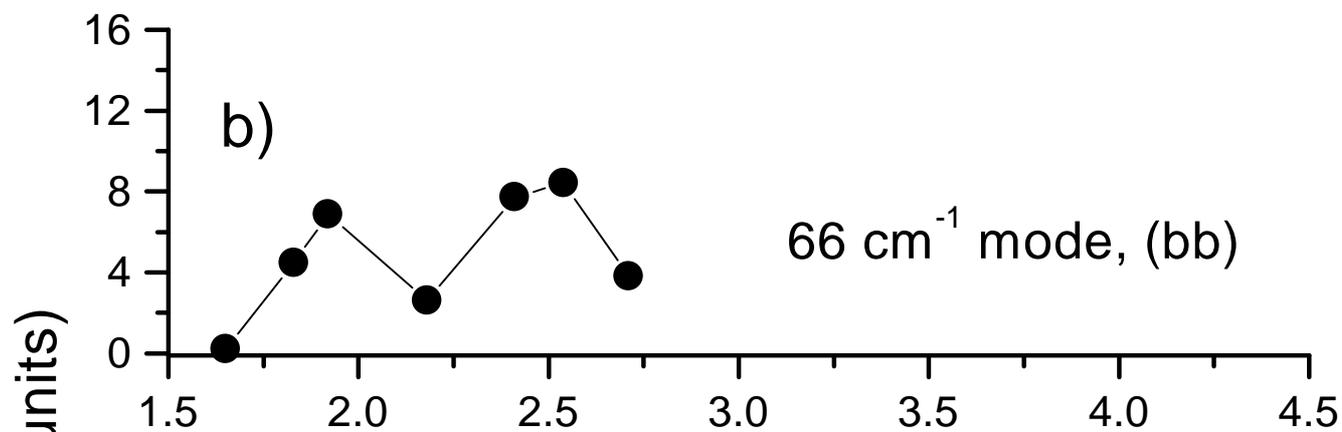
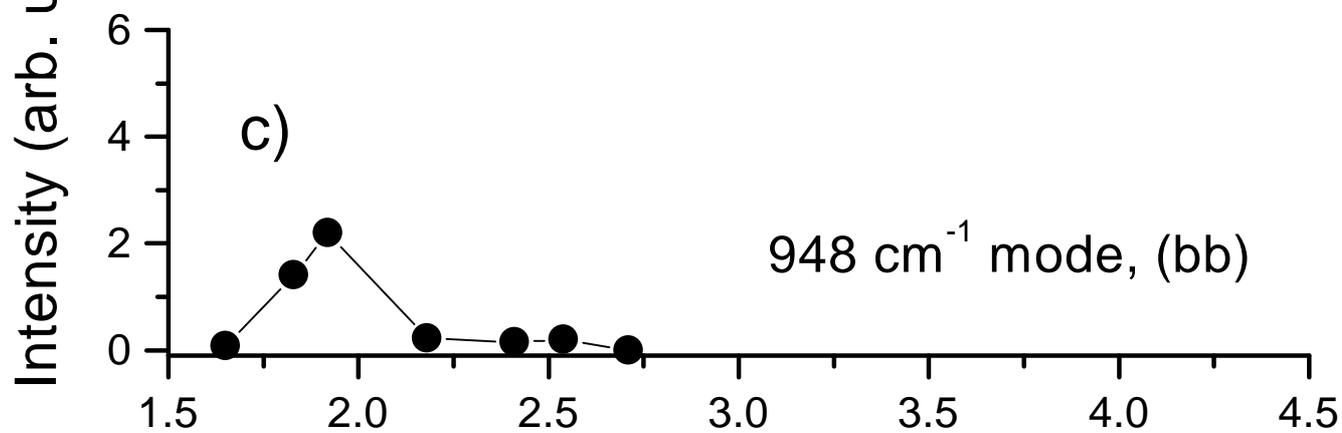
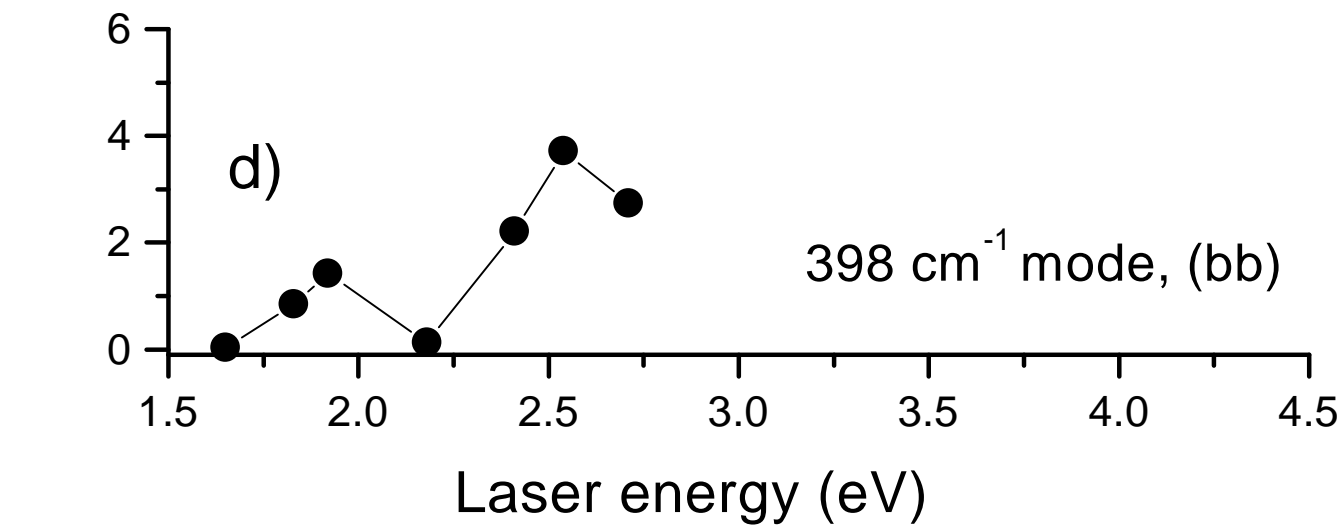
\end{document}